\documentclass[twocolumn,showpacs,floatfix,superscriptaddress]{revtex4}
\usepackage{graphicx}
\usepackage{bm}

\begin{document}

\title{Mean-field analysis of the stability of a K-Rb Fermi-Bose mixture}

\author{M. Modugno}
\affiliation{%
  LENS - Dipartimento di Fisica, Universit\`a di Firenze and INFM\\ Via
  Nello Carrara 1, 50019 Sesto Fiorentino, Italy}
\affiliation{%
BEC-INFM Center, Universit\`a di Trento, 38050 Povo, Italy}
\author{F. Ferlaino}
\author{F. Riboli}
\altaffiliation{%
Present address: 
INFM-Dipartimento di Fisica, Universit\`a di Trento, Via Sommarive 14, 
38050 Povo, Italy}
\author{G. Roati}
\author{G. Modugno}
\author{M. Inguscio}
\affiliation{%
  LENS - Dipartimento di Fisica, Universit\`a di Firenze and INFM\\ Via
  Nello Carrara 1, 50019 Sesto Fiorentino, Italy}

\date{\today}

\begin{abstract}
We compare the experimental stability diagram of a Fermi-Bose
mixture of $^{40}$K and $^{87}$Rb atoms with attractive
interaction to the predictions of a mean-field theoretical model.
We discuss how this comparison can be used to give a better
estimate of the interspecies scattering length, which is currently
known from collisional measurements with larger
uncertainty.
\end{abstract}
\maketitle

The production of degenerate Fermi-Bose mixtures of atoms
\cite{hulet,salomon,ketterle,roati} has opened new directions in
the field of ultracold atomic gases, with relevant implications
for the achievement of fermionic superfluidity. Collisional
interactions between bosons and fermions have a twofold role: they
allow for an efficient sympathetic cooling of the fermions into
quantum degeneracy, and they affect the properties of the
degenerate mixture. In case of a large boson-fermion attraction,
the most important predicted effects are mean-field instabilities
\cite{molmer,roth,miya} and boson-induced interactions between
fermions \cite{induced}, besides a general modification of the
properties of the two individual components \cite{japan,hu,liu}.
We have recently observed the collapse of the Fermi gas in a
$^{40}$K-$^{87}$Rb mixture \cite{science}, which is indeed
characterized by a large attractive interaction between the
components.

In our study of the collapse we found that the order of magnitude
for the critical atom numbers is in agreement with the predictions
of a mean-field model for spherical geometry \cite{roth}.

In this work we investigate the stability diagram of the K-Rb
system, and we make a quantitative comparison of the experimental
findings with mean-field theory taking into account the actual
trapping potential for the atoms. We find that the mean-field
model is able to reproduce the critical atom numbers for collapse
using a value for the boson-fermion scattering length $a_{BF}$
that is in good agreement with the one we determined through the
study of cold collisions \cite{simoni}. Moreover, we show how the
comparison of theory and experiment on the stability diagram can
be used to determine the scattering length with a much lower
uncertainty than that determined from collisional measurements.

The Fermi-Bose mixtures of $^{40}$K (fermions) and $^{87}$Rb (bosons) 
atoms are produced by sympathetic cooling in a magnetic
potential \cite{roati}. Both species are trapped in their
low-field-seeking states with largest magnetic moment, which
experience potentials of the kind
\begin{eqnarray}
V_B({\bf x})&=&\frac{1}{2}m_B\left[\omega_{B\perp}^2(x^2+y^2)+
\omega_{Bz}^2z^2\right]\\    \nonumber
 V_F({\bf
x})&=&\frac{1}{2}m_F\left[\omega_{F\perp}^2(x^2+(y-y_0)^2)+
\omega_{Fz}^2(z-z_0)^2\right]\,.
 \label{eq:trapping}
\end{eqnarray}
The trap frequencies are $\omega_{F\perp}=2\pi\times317$ Hz,
$\omega_{Fz}=2\pi\times24$ Hz for $^{40}$K, while those of
$^{87}$Rb are a factor $\sqrt{m_B/m_F}\simeq1.47$ smaller. The
system is characterized by a gravitational sag between the two
clouds, which affects both the horizontal and vertical directions,
due to a small misalignment of the magnetic trap with respect 
to the direction of gravity \cite{ribo}. The values of the displacement 
between the two
potential centers are $y_0=3.6~\mu$m and $z_0=20~\mu$m.

The largest samples that we are able to cool to quantum degeneracy
amount to about 2$\times$10$^5$ bosons coexisting with
5$\times$10$^4$ fermions. The atom numbers for both species can be
adjusted separately by controlling the loading and the evaporation
procedures. The coldest Fermi gas observed have a
temperature of about 0.2$T_F$, and the minimum temperature of the
bosons that we are able to measure from the non-condensed fraction
is limited to about 0.6$T_c$.

In the mean-field model that we use to describe this Fermi-Bose
system, the ground state is determined by the solution of Gross-Pitaevskii
 (GP) equation for the bosons, coupled to the Thomas-Fermi
equation for the fermions
\cite{roth,molmer}
\begin{eqnarray}
&&\!\!\!\!\!\!\!\!\!\!
\left[-\frac{\hbar^2}{2m_B}\nabla^2 + V_B + g_{BB}n_B + g_{BF}n_F
\right]\phi=\mu_B\phi
\label{eq:bose}\\
&&\!\!\!\!\!\!\!\!\!\!n_F=\frac{(2m_F)}{6\pi^2\hbar^3}^{3/2}
\left(\epsilon_F -V_F-g_{BF}n_B\right)^{3/2}
\label{eq:fermi}
\end{eqnarray}
where $n_B=|\phi^2|$ and $n_F$ are the boson and fermion densities
respectively, and $\phi({\bf x})$  is the condensate wave
function. The boson-boson and boson-fermion interactions are
described by the coupling constants $g_{BB}=4\pi \hbar
^2a_{BB}/m_B$ and $g_{BF}=2\pi \hbar ^2a_{BF}/m_R$ in terms of the
$s$-wave scattering length $a_{BB}$ and $a_{BF}$. Here $m_{B,F}$
are the atomic masses and $m_R=m_B m_F/(m_{B}+m_F)$ is the reduced
mass.

In the case of the rubidium-potassium mixture, the best
experimental estimates of the scattering lengths are
$a_{BB}=98.98\pm 0.02~a_0$ \cite{verhaar} and
$a_{BF}=-410\pm80~a_0$ \cite{simoni}. The difference in the
uncertainties for the two quantities stems from the different
techniques used, which is a combination of photoassociation and
Feshbach spectroscopies in the first case, and measurement of cold
collisions in the second case. Although the investigation of cold
collisions is the easiest way to measure the interactions in an
ultracold gas, it is often accompanied by a rather large
uncertainty. Indeed, what one measures in the experiment is a
collisional rate, which at ultralow temperature has a dependence
\begin{equation}
\Gamma\propto N a^2
 \label{eq:collisions}
\end{equation}
on the atom number $N$ and
on the scattering length $a$. The absolute atom number, in the
case of a mixture of two species, cannot be determined with an
accuracy typically better that 40\%, giving rise to a 20\%
uncertainty on the scattering length.

\begin{figure}
\centerline{\includegraphics[width=7.5cm,clip=]{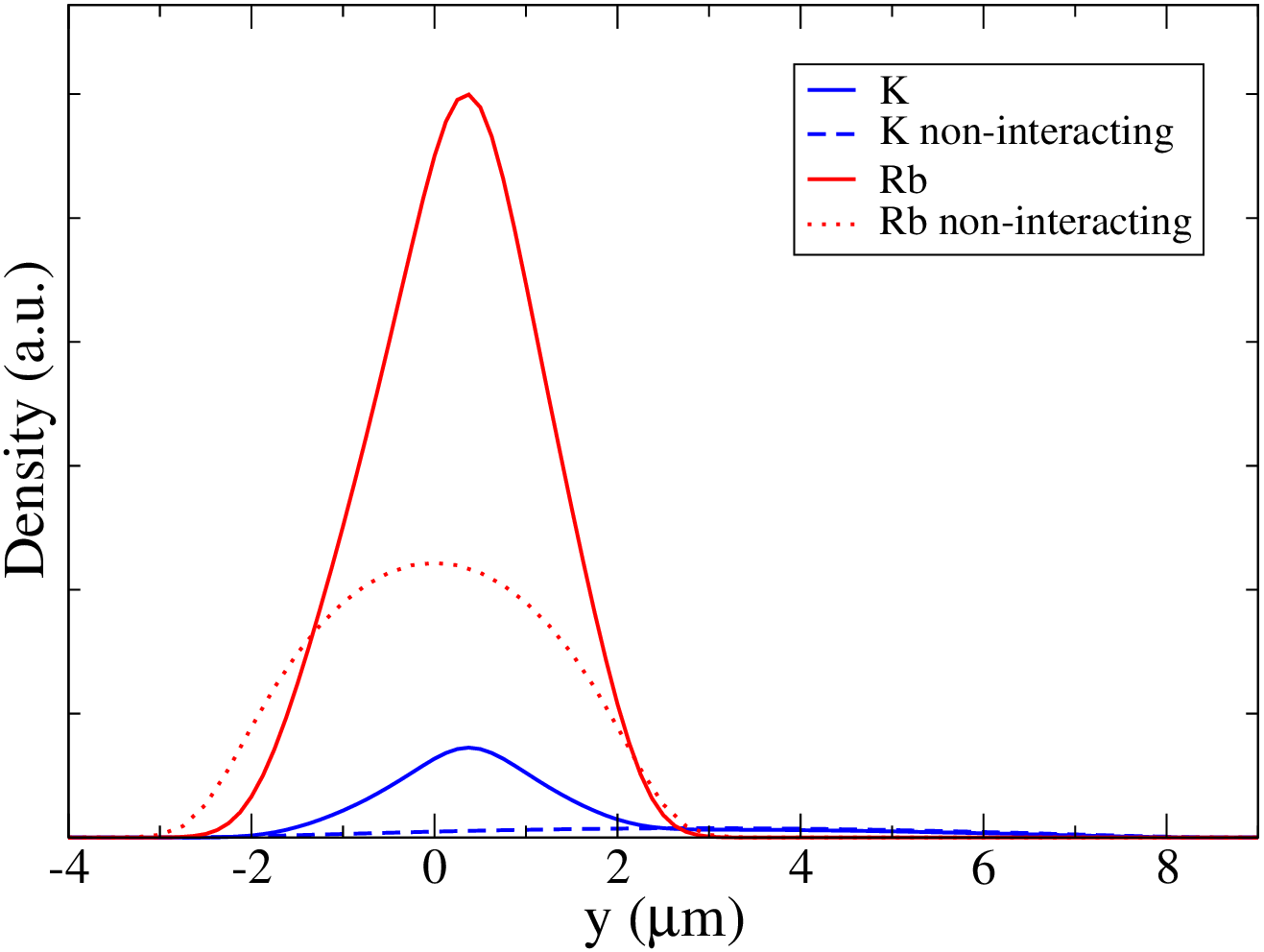}}
\centerline{\includegraphics[width=7.5cm,clip=]{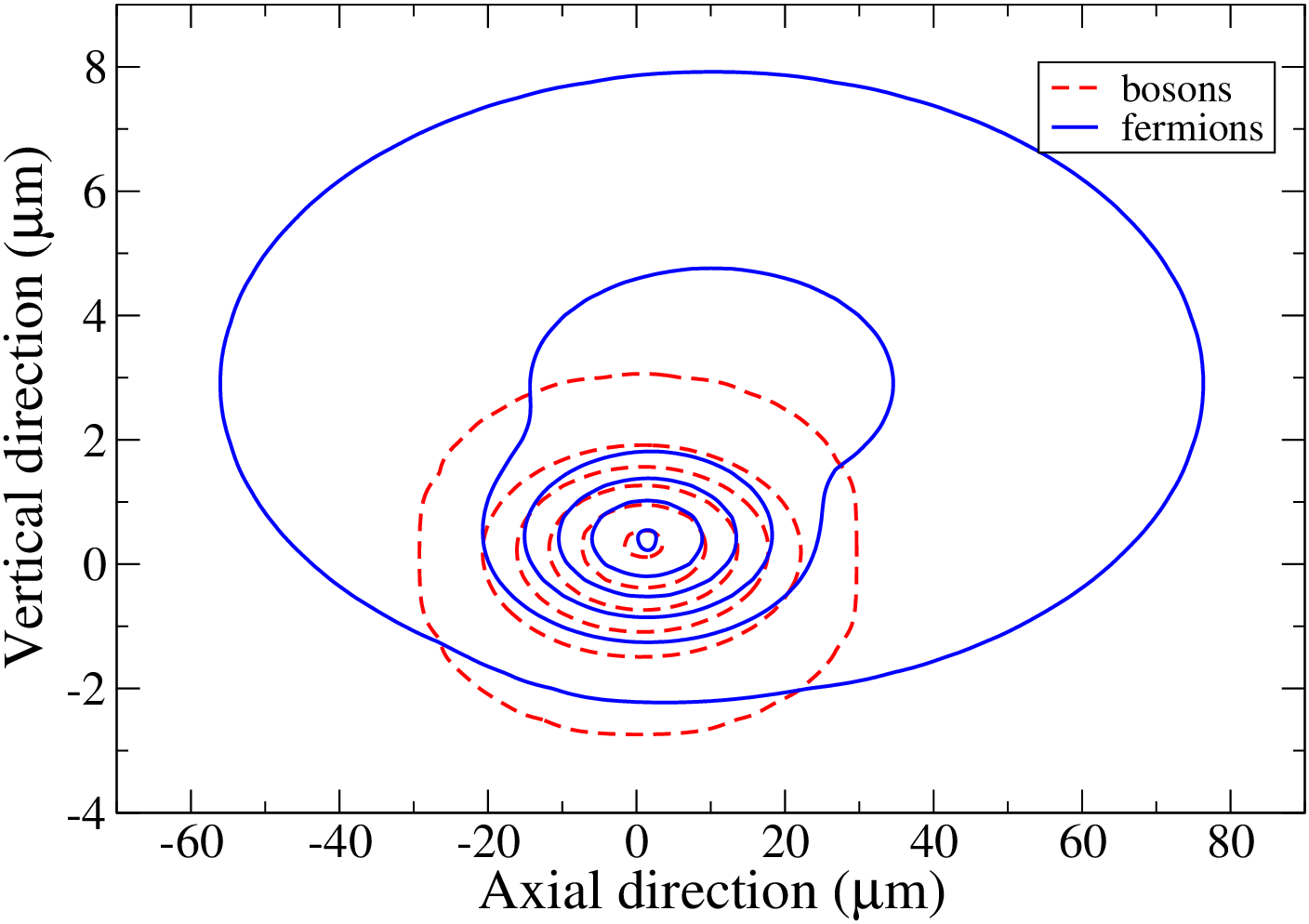}}
\caption{Density profiles along the vertical $y$ direction (a)
and contour levels in the $(y-z)$ plane (b) of the ground-state of
the Fermi-Bose mixture, for
$a_{BF}$=-410~$a_0$, $N_B=7.5\cdot
10^4$, $N_F=2\cdot10^4$.
}
\label{fig:ground}
\end{figure}

By solving equations (\ref{eq:bose})-(\ref{eq:fermi}) we can
investigate the ground state and the stability of the system
against collapse. The equations are solved with an imaginary-time
evolution, embedded in an iterative scheme, as in \protect\cite{roth}.
The stability is checked by requiring that energies, chemical potentials
and central densities converge to the final value, with a precision
at least of $10^{-6}$.
In Fig. \ref{fig:ground} we show a typical ground state configuration,
calculated with the nominal values $a_{BF}$=-410~$a_0$, $N_B=7.5\cdot
10^4$, $N_F=2\cdot10^4$.
As already discussed in Refs. \cite{molmer,roth}, the
mutual attraction results in an enhancement of the density of both
species in the volume of overlap. We note that the gravitational
sag, that shifts the centers of the two clouds, reduces the effect
of the mutual attraction with respect to the concentric case
studied in Ref. \cite{roth}.

The deformed ground state of the system is predicted to lead to a
series of effects that can be described within a mean-field
approach, including a modification of the frequencies of
collective excitations \cite{japan,hu} and of the expansion of the
two clouds from the trap \cite{liu}. We actually observed the
latter effect already from the first experiments \cite{roati},
where we had an evidence of a modified expansion of the condensate
in presence of the Fermi gas.

If the atom numbers are increased above some critical values, an
instability occurs. In our model, accordingly to the study reported in
Refs. \cite{feldmaier,roth} the signature of the instability is
the failure of the convergence procedure during the iterative
evolution toward the ground state of the system. In particular,
the onset of instability is characterized by an indefinite growth
of central densities which triggers the simultaneous collapse of
the two species. We note that the investigation of the actual dynamics after
the system has been driven into the unstable region would require a
description that goes beyond Eqs. (\ref{eq:bose})-(\ref{eq:fermi}),
in a similar fashion to what happens during the collapse of a single
Bose-Einstein condensate with attractive interaction 
\cite{sackett,kagan}.
Here we will not discuss these aspects, since we are concerned with
the determination of the critical values for the onset of instabilities
only.

In the experiment we observe the instability of the Fermi-Bose
system as the number of condensed bosons in increased by
evaporation, at a fixed number of fermions \cite{science}. The
signature of the instability is a sudden loss of most of the
fermions from the magnetic trap, which we attribute to a largely
increased three-body recombination in the collapsing samples. Here
we are interested just in the critical atom numbers, but we note
that further experiments, for example using a Feshbach resonance
to tune $a_{BF}$ \cite{simoni}, will be necessary to study the
collapse dynamics of the system.

\begin{figure}
\centerline{\includegraphics[width=8cm,clip=]{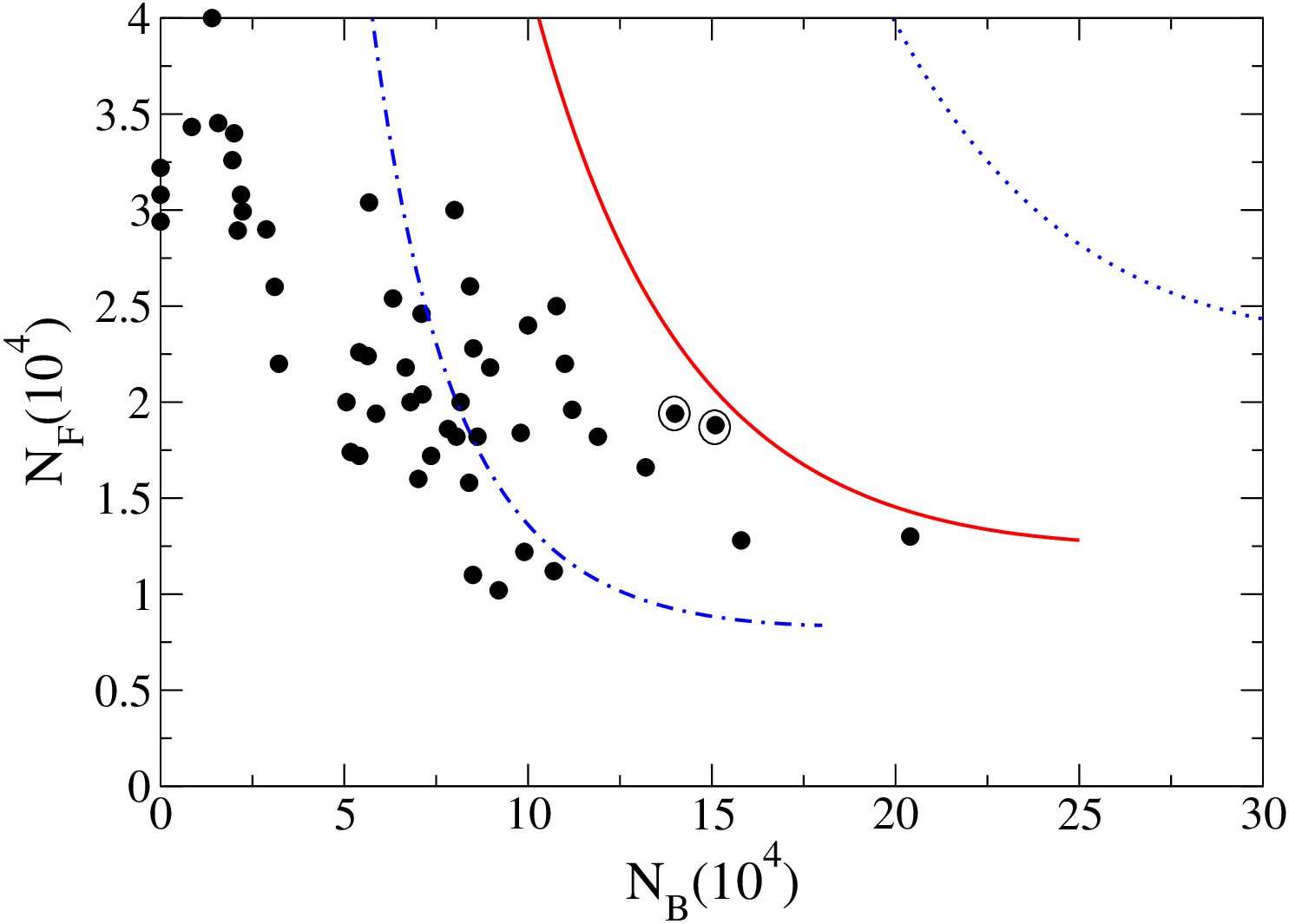}}
\caption{Region of stability of the Fermi-Bose mixture, as a
function of the number of atoms. The black dots are the
experimental points; lines are the theoretical
prediction for the boundary between the stable (left) and collapse
(right) regions, for three values of the inter-particle scattering
length (in units of the Bohr radius $a_0$): $a_{BF}=-380~ a_0$
(dotted line), $a_{BF}=-395~ a_0$ (continuous line), $a_{BF}=-410~
a_0$ (dashed-dotted line). The marked dots are found very close to
the instability (see text).} \label{fig:collapse1}
\end{figure}

To compare the predictions of the mean-field model to the
experimental findings on the instability, we have built a
stability diagram, shown in Fig. \ref{fig:collapse1}. Here we
report in the plane $N_B-N_F$ the condensate and fermion atom
numbers that we were able to measure in the experiment for
\textit{stable} samples, and compare them with the calculated
critical line for instability. We note that all the data points
refer to samples at temperatures of the order of 0.2-0.5$T_F$ for
the Fermi gas, and with almost no detectable thermal fraction for
the BEC. In the model, the occurrence of instability depends on
both $N_B$, $N_F$, and also on the value of the scattering
length $a_{BF}$. In Fig. \ref{fig:collapse1} we have therefore
plotted a family of critical lines by varying $a_{BF}$ around the
nominal value. Note that the position of the critical line depends
quite strongly on the value of $a_{BF}$, and it shifts toward
larger atom numbers for decreasing magnitudes of $a_{BF}$.
Although the critical line calculated for the nominal value
$a_{BF}$=-410 $a_0$ is in rather good agreement with the
experimental observation, the one calculated for $a_{BF}$=-395
$a_0$ better fits the data.  In the experiment we observed the
collapse of the Fermi gas for number pairs close to the two marked
data points \cite{note}, which appears to be actually the region
of achievable number pairs closest to the instability.

\begin{figure}
\centerline{\includegraphics[width=8cm,clip=]{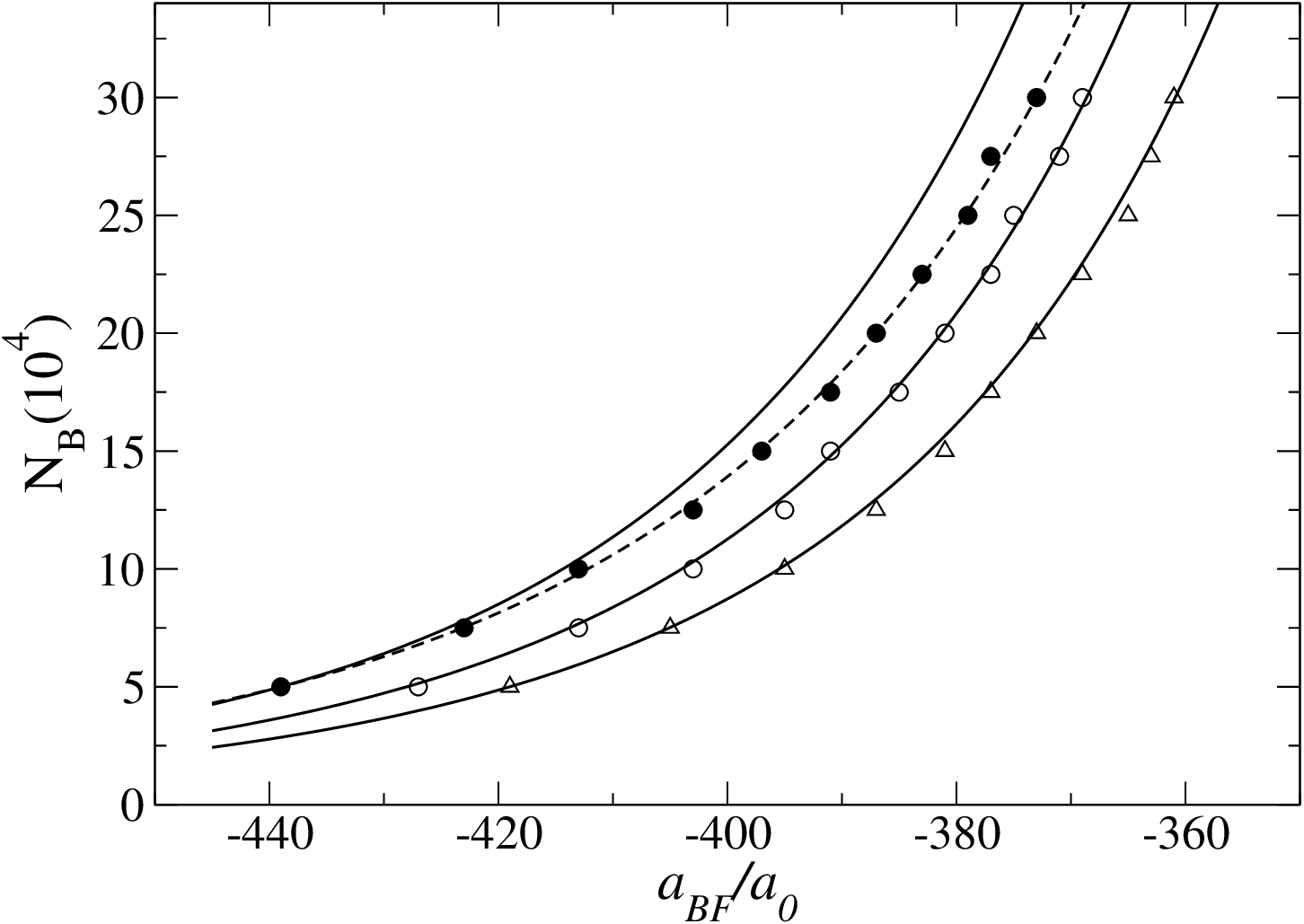}}
\caption{
Behavior of the critical number of atoms $N_B$
as a function of the scattering length $a_{BF}$,
for $N_B=4.25~N_F$ (triangles) and
$N_B=8.5~N_F$ (circles).
The numerical solutions of the mean-field
equations (symbols) are compared with the scaling law in Eq.
(\protect\ref{eq:scaling}) with exponent $\alpha=12$
(continuous line) and $\alpha=11$ (dashed line).
Filled and empty symbols refer respectively
to the full 3D geometry and to the simplified spherically
symmetric case discussed in the text.
}
\label{fig:molmer}
\end{figure}
The strong dependence on the criticality on $a_{BF}$ that we find
in the mean-field analysis is in accordance with the scaling law
for the critical number of condensate atoms
\begin{equation}
N_B^{crit}\sim |a_{BF}|^{-\alpha}
\label{eq:scaling}
\end{equation}
with $\alpha=12$, discussed in \cite{molmer,miya}. Here we have
verified that this scaling behavior perfectly fits the numerical
solutions of Eqs. (\ref{eq:bose})-(\ref{eq:fermi}) in case of a
simplified concentric spherical configuration, by using an average
trapping frequency as usually found in literature
\cite{roth,albus}, as shown in Fig. \ref{fig:molmer}. In this
calculation we have kept the number ratio $N_B$/$N_F$ fixed, which
corresponds to move along a straight line passing from the origin
in the diagram of Fig. \ref{fig:collapse1}. This calculation also
shows that the 3D geometry of the experiment results in a
renormalization of the scaling exponent, $\alpha_{3D}\simeq11$,
besides an effective increase (in modulus) of the critical
scattering length, of the order $2\%$, due to the presence of
gravity.  In fact, the effect of the gravitational shift is a
reduction of the density enhancements due to the attractive
interaction, with respect to the case in which the to species are
trapped by concentric potentials.

\begin{figure}
\centerline{\includegraphics[width=8cm,clip=]{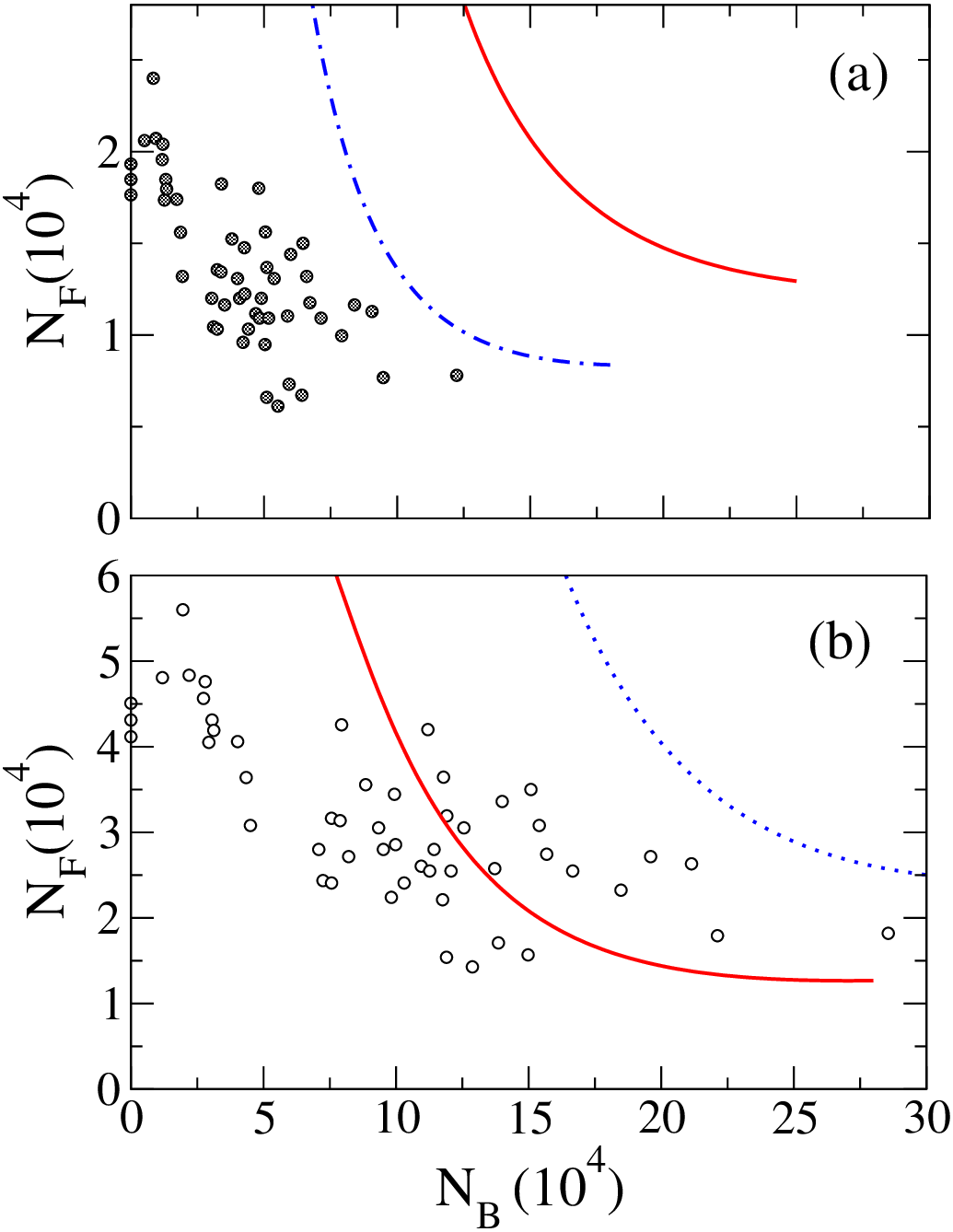}}
\caption{Region of stability of the Fermi-Bose mixture, as a
function of the number of atoms. To keep into account the
experimental uncertainty on the number of atoms these have been
increased or reduced by a $40\%$, respectively in (a) and (b).
Continuous lines are the theoretical prediction for the boundary
between the stable (left) and collapse (right) regions, for three
values of the inter-particle scattering length (in units of the
Bohr radius $a_0$): $a_{BF}=-380~ a_0$ (dotted line),
$a_{BF}=-395~ a_0$ (continuous line), $a_{BF}=-410~ a_0$
(dashed-dotted line).} \label{fig:collapse2}
\end{figure}

This analysis suggests the possibility of using a
comparison of the experimental and calculated critical atom
numbers to extract a precise information on the value of
the scattering length for our K-Rb mixture. To account for the
uncertainty in the determination of atom numbers in the
experiment, in Fig. \ref{fig:collapse2} we have plotted the same
data points of Fig. \ref{fig:collapse1}, expanding or compressing
both scales by 40\%, and compared them with the same family of
critical lines.

By taking the critical lines that
best fit with the observation in these two limiting cases, we can
extract the following mean-field estimate for $a_{BF}$
\begin{equation}
a^{mf}_{BF}\simeq-395\pm15~a_0.
\label{eq:critical}
\end{equation}
Note how the dependence of $a_{BF}$ on the number of atoms of Eq.
\ref{eq:scaling}, stronger than that of Eq. \ref{eq:collisions},
results in a 5-fold smaller uncertainty on the scattering length
determined by the stability analysis than that obtained with
collisional measurements. We expect this
mean-field prediction for $a_{BF}$ in Eq. (\ref{eq:critical}) to
be rather robust, since the $40\%$
experimental uncertainty on the number of atoms
should be quite larger that the possible discrepancy between the GP
theory and the actual behavior of the system, as recently discussed
for the case of a single $^{85}$Rb Bose-Einstein condensate
with attractive interaction \cite{bradley,report,savage}.

The assumption of the same scaling for boson and fermion atom
numbers due to experimental uncertainty is not unrealistic.
Indeed, the possible sources of systematic errors in the atom
number calibration are residual magnetic fields, non perfect
polarization of the probe beam, or improper calibration of the
imaging magnification. When using a simultaneous imaging scheme
\cite{roati}, all these factors affect in the same way the
measured atom numbers for bosons and fermions.

We note that recently it has been proposed a beyond mean-field
approach, including  the second order correction in the 
scattering length $a_{BF}$, that should
provide important modifications to the mean-field results
\cite{albus}. Nonetheless we have verified that the inclusion of
this term stabilizes the system in the range of atom numbers
considered here for any value of the scattering length, thus
forbidding the collapse, in contrast with the experimental
findings. Higher order terms might therefore be important to
provide a correct beyond mean-field description, although at this
stage it is not clear whether this could affect the critical
number of atoms for collapse.

In conclusion, our mean-field analysis of the stability of the
K-Rb Fermi-Bose mixture shows the possibility of an independent
determination of the relevant scattering length of the system. The
strong dependence of the critical atom numbers on the scattering
length allows to largely reduce the effect of the experimental
uncertainty. The value we determine for $a_{BF}$ is in very good
accordance with that found from collisional measurements. We note
that an independent, more precise determination of $a_{BF}$, for
example using Feshbach spectroscopy \cite{simoni}, would be useful
to test the validity of the mean-field theory for strongly
interacting Fermi-Bose mixtures, and to asses the importance of
beyond mean-field effects.

\begin{acknowledgments}
This work was supported by MIUR, by EC under contract
HPRICT1999-00111, and by INFM, PRA ``Photonmatter''.
We acknowledge useful discussions with A. Albus and F. Illuminati.
\end{acknowledgments}


\begin{thebibliography}{}
\bibitem{note} Those two points represent the last stable samples
along to two evaporation routes
that led to the collapse, in the series of measurements described
in \cite{science}.
\end{thebibliography}
\end{document}